# A deep learning approach for brain tumor detection using magnetic resonance imaging

Al-Akhir Nayan[1], Ahamad Nokib Mozumder[2], Md. Rakibul Haque[2], Fahim Hossain Sifat[3], Khan Raqib Mahmud[3], Abul Kalam Al Azad[3], Muhammad Golam Kibria[3]
[1]Department of Computer Engineering, Faculty of Engineering, Chulalongkorn University, Bangkok, Thailand
[2] Department of Computer Science and Engineering, Faculty of Sciences and Engineering, European University of Bangladesh, Dhaka, Bangladesh
[3]Department of Computer Science and Engineering, School of Science and Engineering, University of Liberal Arts Bangladesh, Dhaka, Bangladesh

**ABSTRACT**

The growth of abnormal cells in the brain's tissue causes brain tumors. Brain tumors are considered one of the most dangerous disorders in children and adults. It develops quickly, and the patient's survival prospects are slim if not appropriately treated. Proper treatment planning and precise diagnoses are essential to improving a patient's life expectancy. Brain tumors are mainly diagnosed using magnetic resonance imaging (MRI). As part of a convolution neural network (CNN)-based illustration, an architecture containing five convolution layers, five max-pooling layers, a Flatten layer, and two dense layers has been proposed for detecting brain tumors from MRI images. The proposed model includes an automatic feature extractor, modified hidden layer architecture, and activation function. Several test cases were performed, and the proposed model achieved 98.6% accuracy and 97.8% precision score with a low cross-entropy rate. Compared with other approaches such as adjacent feature propagation network (AFPNet), mask region-based CNN (mask RCNN), YOLOv5, and Fourier CNN (FCNN), the proposed model has performed better in detecting brain tumors.

*Corresponding Author:*

Al-Akhir Nayan
Department of Computer Engineering, Faculty of Engineering, Chulalongkorn University
Bangkok, Thailand
Email: asquiren@gmail.com

## 1. INTRODUCTION

The human body is made of millions of cells, and the brain is an essential part of the human body. A brain tumor is caused by abnormal cells in the brain's tissue. It is considered one of the world's deadliest diseases [1], [2] due to its escalating impact and fatality rate in all age categories. It is India's second-leading cause of cancer [3]. According to the American Cancer Society's most recent report, around 24,000 people in the United States were infected with brain tumors in 2020, with an estimated 19,000 deaths due to the increased use of technology such as cell phones and tablets [4], [5]. Approximately 120 varieties of tumors have been discovered too far, and they all arise in various shapes and sizes, making diagnosis more challenging [6]–[8]. Medical imaging modalities like positron emission tomography (PET), computed tomography (CT), magnetic resonance imaging (MRI), and magnetoencephalography (MEG) have been utilized to diagnose brain irregularities for a long time [9], [10]. The MRI multimodality imaging technique is the most common and efficient technology routinely used to diagnose brain tumors because of its capacity to distinguish between structure and tissue based on contrast levels [11], [12]. MRI anomaly detection is now primarily manual, and doctors must spend significant time discovering and segmenting the tumor for therapy and surgery [13], [14]. This manual procedure is also prone to errors and can endanger one's life. Researchers have begun to examine various machine learning and deep learning techniques for computer-based tumor diagnosis and segmentation to address these challenges.

Deep learning is a machine learning subfield widely used to develop a semi-automatic, automatic, or hybrid model to detect and segment tumors in less time [15]. Radiologists can make a more accurate prognosis and increase the odds of long-term survival if a brain tumor is discovered early [10]. The tumor's shifting appearance, position, form, and size [16] remains a complex process. There has already been a lot of work done to assist doctors, patients, and researchers. Many computers aided diagnostic (CAD) systems have been created to detect and classify brain anomalies [17] automatically, but they still perform with poor accuracy [18]. Several articles have been published without highlighting the flaws in previous work or providing any significant insight into future directions. Interoperability is lacking in most hybrid models, while gradient vanishing is a concern in deep learning models. Similarly, there is a lack of uniformity in data preprocessing.

This article aims to process the images from MRI and detect tumors in the brain by solving previous issues. Image enhancement, rebuilding, and estimation extraction techniques have been applied to enhance image quality while preparing the dataset. The image digitization process and picture upgrade techniques handle defective images. A modified CNN has been applied to take an MRI scan image of the brain as input, detect the tumor, and give the result as output. The network contains five convolution layers, five max-pooling layers, a Flatten layer, multiple hidden layers, and two dense layers. A modified recto-triangular architecture has been utilized in the hidden layer that enhances the probability distribution. The model's accuracy has been evaluated and compared with state-of-the-art techniques. The model has performed better than other approaches in detecting brain tumors from MRI images.

## 2. RELATED WORKS

Image detection plays a crucial role in analyzing brain tumors using MRI images. Many methods for detecting brain tumors from MRI images have been proposed. In the method presented by Kumar *et al*. [19], brain tumors were predicted using a fully convolutional neural network (FCN) from MRI images.

Derikvand *et al*. [20] presented an approach based on neural convolution. The method used glioma brain tumor detection networks in MR imaging. The proposed process was a hybrid, multiple CNN architectures using local and global brain tissue knowledge to predict each pixel's label, improving results.

CNN was used for image segmentation and detection by Hemanth *et al*. [21]. It explicitly pulled features from images with the least amount of preprocessing. LinkNet was employed. The architecture of a neural network was designed to conduct semantic segmentation and detection. The LinkNet network blocks of encoders and decoders were responsible for breaking and rebuilding the image until it was routed through a few final levels of convolution. root mean square error (RMSE), recall, sensitivity, precision, F-score specificity, and percentage mean error (PME) evaluated the suggested CNN's performance.

Hossain's methodology [22] began with the input image's skull striping, which removed the skull part from the MRI images. The fuzzy C-means clustering algorithm was used to detect the filtered image. Texture-based and statistical-based features were extracted from the images. The extracted features were fed into a CNN model. The model's accuracy for brain tumor prediction was high, but the computational time was much longer than other models.

Minz *et al*. [23] proposed one method that extracted characteristics using the gray level co-occurrence matrix methodology. An image's texture was defined by calculating specified spatial relationships that appeared in an image. Gray level co-occurrence matrix was built, and this matrix extracted statistical measurements. The classification was done with the AdaBoost classifier, and the proposed system attained an accuracy of 89.90%.

Gurbin *et al*. [24] proposed a method using discrete wavelet transform levels (DWT) and continuous wavelet transform levels (CWT). Support vector machines (SVM) were utilized to identify benign, malignant, or healthy brains. The research suggested that CWT performs better than DWT in computation.

Chander *et al*. [25] proposed a strategy to detect afflicted brain tissues using the grade 4 gradient boosting machine (GBM). The retrieved properties of Bayesian naive were used to classify them. This method yielded an accuracy of 83.33%. The MRI image was initially fed into the algorithm, which filtered it to smooth it out and remove noise. The next step was to mask the filtered image to remove brain tissues from the skull. Finally, the acquired features were sent into an SVM classifier, which determined whether the image was malignant or not.

After examining many algorithms for detecting brain tumors, we found that most algorithms required a long computational time, while cheap computational time methods could not provide greater accuracy. Every classifier requires rectifying the predecessor's flaws, and boosting is sensitive to outliers. As a result, the technique is overly reliant on outliers. Another downside is that scaling up the process is nearly impossible because each estimator is based on the accuracy of preceding predictors. As a result, we intended to propose an approach that could provide excellent accuracy.

## 3. BRAIN TUMOR DETECTION PROCESS
### 3.1. Data collection
A dataset containing 30,000 images was collected to train the deep learning model. The dataset contained two classes: fresh MRI images of the brain and the images with a brain tumor. The collected data contains 15,000 images of healthy brains and 15,000 images of brain tumors. For testing the dataset, related images were collected from Google. There are 4,400 brain images that were taken without a tumor, and 3,200 images were taken with a tumor. A small part of the collected images is shown in Figure 1. For comparison purpose, the BRATS datasets [26] *(BRATS_2018, BRATS_2019, BRATS_2020)* were utilized.

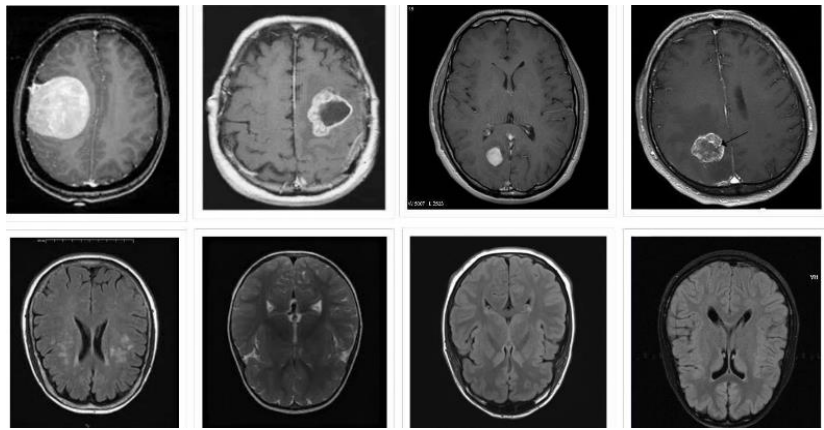

Figure 1. A part of collected images

### 3.2. Data preprocessing
The preprocessing step aims to increase picture quality, data cleansing, and contrast-enhancing. The median filter filters out the noise and retrieves valuable data. Median filtering is a nonlinear approach used to retain sharp features in MRI images. In this work, an MRI image was preprocessed by converting the picture to greyscale and using a 3×3 median filter to eliminate noise, which enhanced image quality using (1).

$$f(x,y) = median_{(s,t)eSxy}\{g(s,t)\} \tag{1}$$

A high pass filter was used to locate edges in the acquired MRI image. The edge-identified MRI image was combined with the original image to obtain the enhanced image. The dataset was enlarged using the data augmentation process to keep a strategic distance from overfitting. The dataset was augmented using four distinct tactics: rotate left -90 degrees, rotate left -180 degrees, rotate left -270 degree, and flip every image once.

### 3.3. CNN model architecture
CNN was utilized to detect brain cancers using MR images in this study. CNN is an artificial neural network (ANN) designed to analyze image pixels and extract meaningful images. CNNs are used in image and video recognition [27]–[29], natural language processing, and artificial intelligence. In this study, the proposed architecture contains an input layer, five convolutions, five max pooling, one flattens, fully connected, or concealed, and two dense layers. The basic architecture of the proposed CNN is depicted in Figure 2.

The input layer is usually a pixel-filled image, and to construct a convolution layer, a feature map is built and slid over these pixels. The pooling stage minimizes the number of features and increases the correlation between proximity pixels. The suggested method uses the max-pooling methodology to downsample images and extract essential features, such as edges. In this article, a max-pooling approach is applied after each convolution layer. In the convolution 2D layer, the input images are scaled to 300*300 pixels. The rectified linear unit (ReLU) activation function is used in each convolution layer. The first convolution layer uses sixteen feature maps or filters and a 3×3 feature detector matrix. With the same 3×3 feature detector matrix size, the number of feature maps or filters in the second convolution layer was increased to 32. The max-pooling layer uses a 2×2 feature extraction matrix after the convolution layer. After all the convolutions and max pooling, the resultant matrices are broken down into a flatten layer or a single column with all the pixel values. This flattening layer is then utilized to input the hidden layer of the following artificial neural network. Hidden layers are followed by two dense layers, which process the output for the network. This research compares the assessment score of the proposed recto-triangular design for the hidden layer to triangular and rectangular architectures shown in Figures 3(a) to (c).

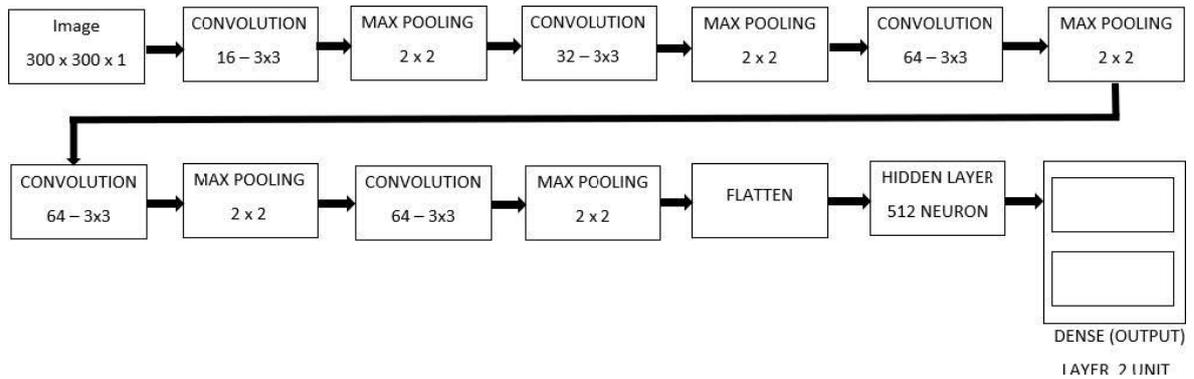

Figure 2. CNN model architecture

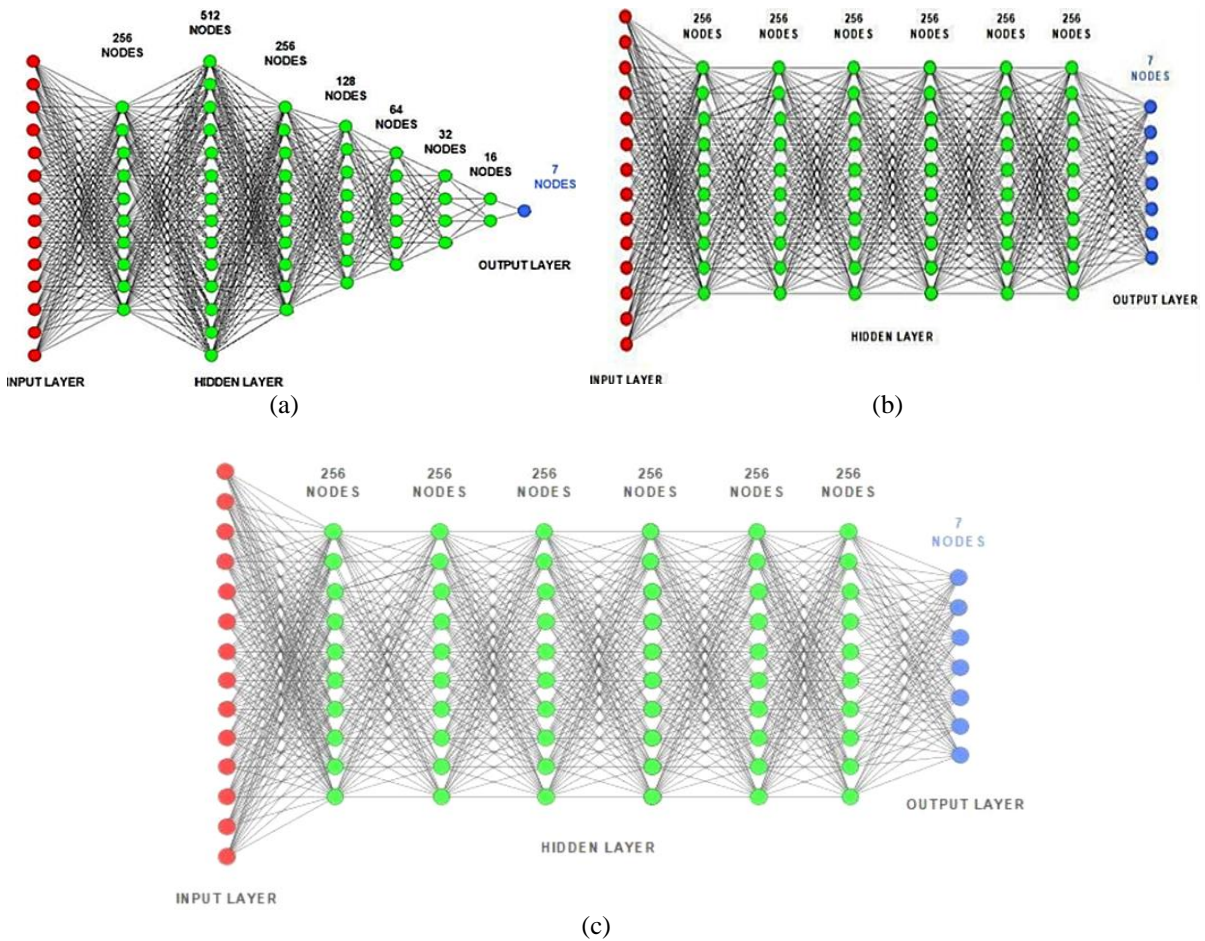

Figure 3. Hidden layer architecture (a) triangular, (b) rectangular, and (c) recto-triangular

### 3.4. Triangular architecture

The modified triangular architecture comprises 256 nodes in the first hidden layer, 512 nodes in the second hidden layer, and the number of nodes drops to take the shape of a triangle from the third layer to the seventh layer. The hidden layers include 256, 512, 256, 128, 64, 32, and 16 nodes. The ReLU activation function activates all the seven hidden levels. The output layer employs the SoftMax activation function to simplify modeling probability distributions. Figure 3(a) depicts the architecture.

### 3.5. Rectangular architecture

There are six hidden layers in the rectangular architecture. Because of the shape, the design is rectangular architecture, which comprises six levels with similar nodes in individual layers. Figure 3(b) illustrates the architecture. The number of hidden layers employed in this article for the rectangular design is 6, with each layer having 256 nodes. Each hidden layer utilizes the ReLU activation function, but only the output layer uses the SoftMax activation function.

### 3.6. Proposed recto-triangular architecture

The recto-triangular, a combination of rectangular and triangular architecture, is proposed in this article. The architecture is made up of six layers in the hidden layer. Figure 3(c) depicts the proposed recto-triangular architecture's shape and layers. The first, second, third, fourth, fifth, and sixth hidden layers contain 512, 256, 128, 128, 256, and 512 nodes. The number of nodes in the prevalent structure of the hidden layer drops at first, then rises till it reaches the output layer. ReLU is employed to activate all six hidden layers. The output layer uses a SoftMax activation function to represent the probability allocation accurately.

## 4. RESULT AND ANALYSIS
### 4.1. Training the model

For evaluating the performance of the training period, a cross-validation approach was used in training. The data was trained using two separate approaches. The first technique separated the data into ten equivalent areas to ensure that each part was equally available. Another strategy was employed to divide the data into ten equal portions, each containing only data from one participant. As a result, the individual package contained data from multiple subjects regardless of the brain tumor class selected by subject-wise cross-validation. This method evaluated a network's ability to generalize medical diagnoses. In clinical practice, the capacity for generalization indicates that a diagnosis can be predicted based on information acquired from participants with no results during training. The focal loss function (2) was operated to tackle class inequality issues.

$$Focal = -(1-P)^\gamma \sum_{n-1}^{n} l_n * \ln(P_n) \dots \qquad (2)$$

The focal loss was expressed as pixel weights, where n is the number of classes, indicating that the pixels belong to the kth class, $P_n$ denotes the anticipated probability, and p denotes a high probability that is more difficult to identify accurately. The focus loss function value is ten, and weights are assigned to classify pixels effectively.

### 4.2. Performance metrics

With the validation findings and four assessment parameters, we created our model. Correct values are true negative (TN) and true positive (TP), where TP indicates accurately classified aberrant brain images and TN indicates accurately classified standard brain images. False-negative (FN) and false-positive (FP) are inaccurate classifications, where FP indicates inaccurate typical brain imaging and FN indicates inaccurate pathological brain images. Our suggested model's accuracy, dice score/F1, recall, and precision is evaluated using (3) to (6).

$$Accuracy = \frac{TP+TN}{TP+FP+FN+TN} \qquad (3)$$

$$Precision = \frac{TP}{TP+FP} \qquad (4)$$

$$Recall = \frac{TP}{TP+FN} \qquad (5)$$

$$F1 = \frac{2*Recall*Precision}{Recall+Precision} \qquad (6)$$

### 4.3. Performance measurement

The MRI data were separated into validation, testing, and training. During training, the suggested framework utilized a minibatch of size 16. An 'Adam' optimizer with a learning rate of 0.001 was used, and the data was shuffled in each epoch. The performance was assessed using four performance matrices (accuracy, precision, dice score, and recall). The suggested model took 31.53 minutes to train, with 47.82 seconds average training time per epoch.

Figure 4 shows training and validation loss and accuracy, whereas 4(a) depicts the training and validation accuracy with the increasing number of epochs. Figure 4(b) depicts the loss curve. From the figure, it is noticeable that the network quickly began learning from MRI images. From a test dataset of 7,600 images, we calculated precision, recall, and F1-score. Some resulting images of tumor detection by the proposed method are shown in Figure 5.

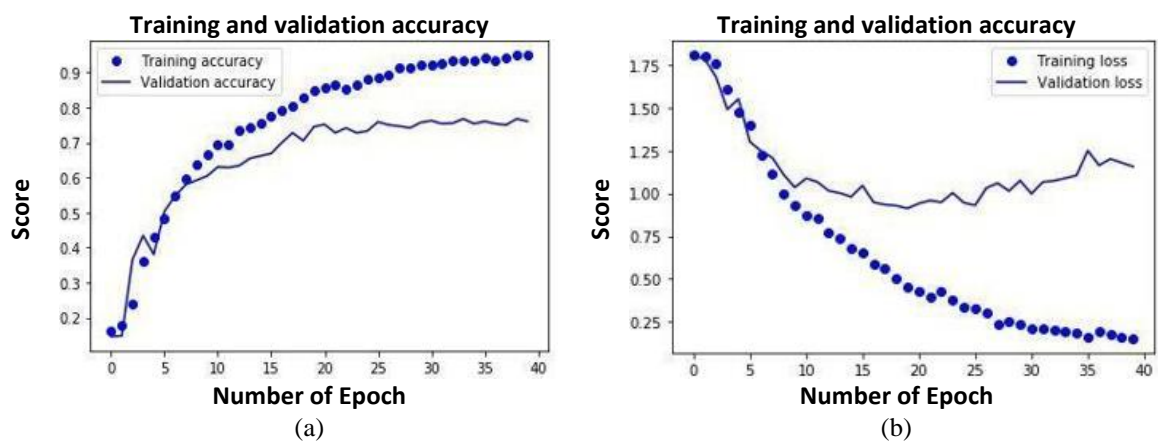

Figure 4. Train vs. validation: (a) accuracy and (b) loss

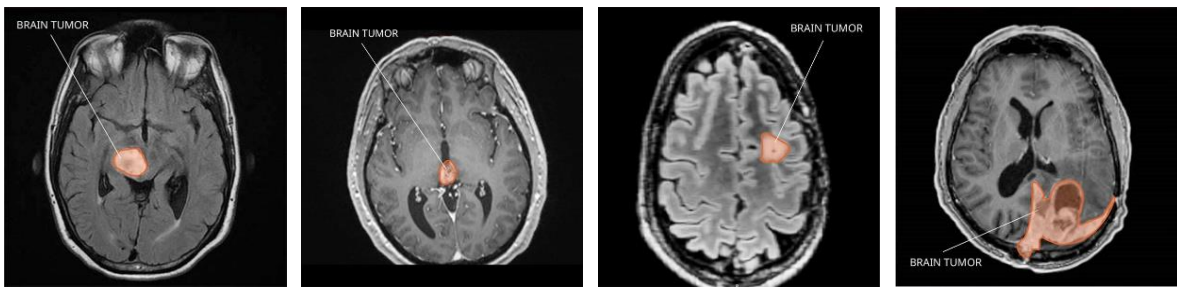

Figure 5. Tumor detection by the proposed model

### 4.4. Comparison among architectures

Triangular, rectangular, and recto-triangular architecture are covered in this study. The same dataset was used to train and evaluate all three architectures. The rectangular architecture, which has six hidden layers, performed a training accuracy of 97.9% and a precision of 91.2%. In contrast, the triangular architecture with seven hidden layers performed a training accuracy of 97.5%, which was 0.4% less than the rectangular architecture. Still, the triangular architecture's precision score was 2.6% greater than the rectangular architecture. Figure 6 shows the comparison of three different architectures. Figure 6 shows that the proposed recto-triangular architecture has achieved a 98.6% training and 97.8% precision score, which is the best compared to the other two. From the evaluation, it can be concluded that the proposed architecture provides satisfactory results and performs better in detecting brain tumors.

### 4.5. Comparison among several approaches and datasets

The suggested model was compared to some existing approaches in terms of performance. As a baseline, the suggested technique was compared to FCNN, mask RCNN, YOLOv5, and AFPNet. Based on the prepared dataset, Table 1 shows a detailed comparison. The findings show that the proposed architecture performs much better than prior research investigations. The generated model dominates and surpasses the current state-of-the-art model. The proposed model's performance was measured on the existing brain tumor dataset BRATS 2018, 2019, and 2020. Table 2 shows the model's performance on BRATS datasets.

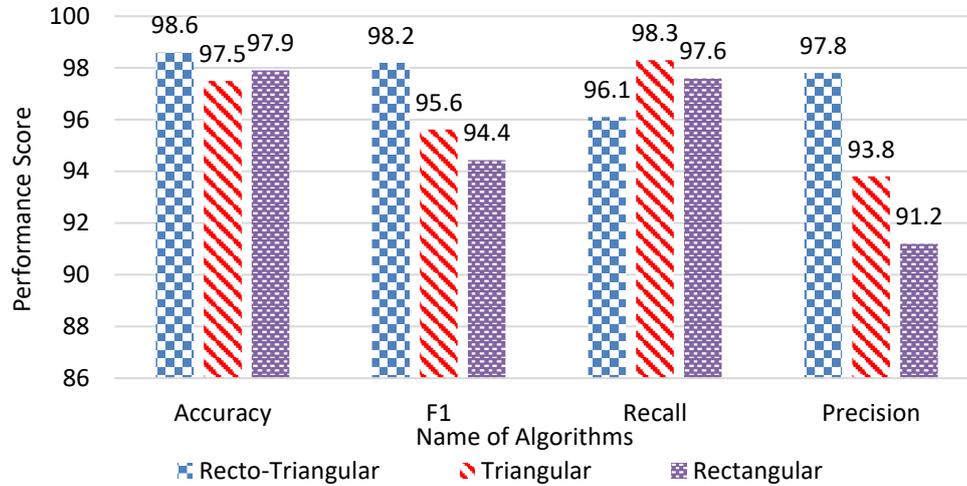

Figure 6. Comparison among architectures

Table 1. Performance comparison of several approaches

| Name | Accuracy | F1 | Recall | Precision |
|---|---|---|---|---|
| AFPNet | 98.30 | 92.67 | 98.25 | 87.65 |
| Mask RCNN | 99.16 | 91.09 | 99.40 | 84.24 |
| YOLOv5 | 98.12 | 93.89 | 97.81 | 90.01 |
| FCNN | 98.73 | 95.98 | 98.50 | 93.37 |

Table 2. Proposed model's performance on BRATS dataset

| Name | Accuracy | F1 | Recall | Precision |
|---|---|---|---|---|
| BRATS_2018 | 97.25 | 96.32 | 96.17 | 95.40 |
| BRATS_2019 | 98.10 | 97.63 | 96.30 | 97.16 |
| BRATS_2020 | 98.49 | 98.21 | 96.58 | 97.53 |

## 5. CONCLUSION

A modified architecture has been proposed in this article that takes advantage of the processed MRI dataset and proposed recto-triangular architecture in the hidden layer for brain tumor detection. The proposed CNN model may perform better than human observers by focusing on a portion of the brain image near the tumor tissue. The proposed preprocessing techniques remove many irrelevant pixels from the images to reduce computing time and capacity. Compared to state-of-the-art alternatives, the proposed model with the proposed hidden layer architecture and the processed dataset have performed better. In the future, we plan to improve the filters to enhance accuracy.

## BIOGRAPHIES OF AUTHORS

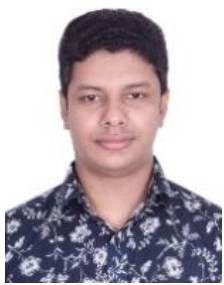 **Al-Akhir Nayan** received a Bachelor of Science degree in Computer Science and Engineering from the University of Liberal Arts Bangladesh (ULAB), Dhaka, Bangladesh, in 2019. He joined the European University of Bangladesh (EUB), Dhaka, in 2019 and worked as a Lecturer in Computer Science and Engineering Department. He is pursuing a master's degree with the Department of Computer Engineering, Chulalongkorn University, Bangkok, Thailand. His research interests include deep learning, machine learning, artificial intelligence, medical image processing, and IoT. He can be contacted via email asquiren@gmail.com.

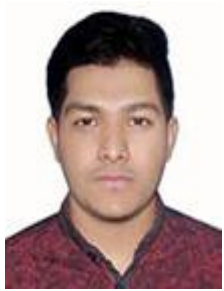

**Ahamad Nokib Mozumder** 🆔 🅶 SC 🔄 received his undergraduate degree from University of Asia Pacific, Dhaka, Bangladesh. He is currently a lecturer at the European University of Bangladesh, Dhaka, Bangladesh. He does research in machine learning, deep learning, and computer vision. He can be contacted via email nokib016@gmail.com.

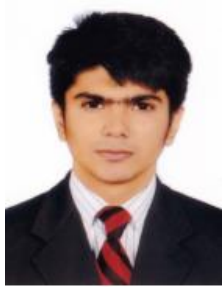

**Md. Rakibul Haque** 🆔 🅶 SC 🔄 received a Bachelor of Science degree in Computer Science and Engineering from the European University of Bangladesh (EUB), Dhaka, Bangladesh. He does research in deep learning, computer vision, and IoT. He can be contacted using e-mail: remon.rakibul.star@gmail.com.

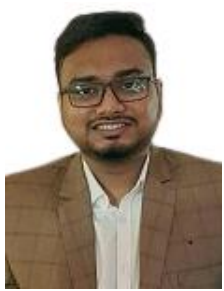

**Fahim Hossain Sifat** 🆔 🅶 SC 🔄 received a Bachelor of Science degree in Computer Science and Engineering from the University of Liberal Arts Bangladesh (ULAB), Dhaka, Bangladesh. He does research in machine learning and Internet of things. He can be contacted via email fahimsifat29@gmail.com.

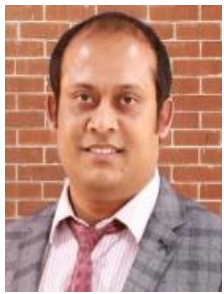

**Khan Raqib Mahmud** 🆔 🅶 SC 🔄 received a double master's in science degree in Computer Simulation for Science and Engineering and Computational Engineering from Germany and Sweden. He is currently working as a lecturer within the Department of Computer Science and Engineering at the University of Liberal Arts Bangladesh (ULAB). He does research in image analysis and computer vision, machine learning, computational aerodynamics, high end particle simulation, and modeling. He can be contacted via email raqib.mahmud@ulab.edu.bd.

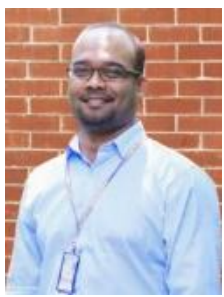

**Abul Kalam Al Azad** 🆔 🅶 SC 🔄 received a Ph.D. in Applied Mathematics from the University of Exeter, UK. He is an Associate Professor and Acting Head of the Computer Science & Engineering (CSE) department at the University of Liberal Arts Bangladesh (ULAB). He does research in data analysis, deep learning, artificial neural networks, and fluid dynamics. He can be contacted via email abul.azad@ulab.edu.bd.

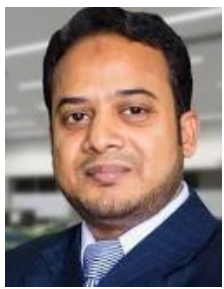

**Muhammad Golam Kibria** 🆔 🅶 SC 🔄 received his Ph.D. degree from the Department of Computer and Information Communications Engineering (CICE) at Hankuk University of Foreign Studies in Korea. His research interests include the Internet of things (IoT), semantic web, ontology, and web of objects (WoO). He can be contacted via golam.kibria@ulab.edu.bd.